# HIGH FORCE DENSITY MULTI-STAGE ELECTROHYDRODYNAMIC JETS USING FOLDED LASER MICROFABRICATED ELECTRODES

*Daniel S. Drew and Sean Follmer*
Stanford University, California, USA

## ABSTRACT

The electrohydrodynamic (EHD) force produced by ions ejected from a corona plasma is a solid state, silent mechanism for accelerating air, useful for applications ranging from electronics cooling to flying microrobots. This paper presents the theoretical motivation and the first implementation of a multi-stage, highly miniaturized EHD device, which can provide both improved absolute power output and power density as compared to single-stage devices. A laser microfabricated, folded electrode design reduces component count and assembly time. Data from one, two, and three-stage devices demonstrates a near linear scaling of output force with stage count, indicating inter-stage ducting successfully reduces losses. Device lifetime is assessed to validate the use of stainless-steel emission electrodes. Areal thrust, force density, and volumetric power density for the three-stage device are among the highest ever measured from an EHD actuator.

## INTRODUCTION

Corona discharge-based electrohydrodynamic (EHD) fluid acceleration is a mechanism by which an applied dc or ac voltage can be converted directly to mechanical work in the form of an induced fluid velocity. It is suitable for applications ranging from electronic device cooling to micro aerial vehicle propulsion due to its solid-state nature and amenability to miniaturization [1]. Recent efforts utilizing EHD for fixed wing flight have incorporated large electrode gap thrusters, yielding similar *areal* thrust densities to smaller gap devices, but relatively low *volumetric* thrust densities [2]. Instead, we propose that stacking small-gap acceleration stages in series has the potential to increase the overall system power density, allowing for a higher efficiency device operating point without dramatically increasing system size.

This paper details the initial efforts to explore the space of serially integrated (i.e., successive acceleration stage) millimeter-scale EHD devices. We illustrate how this approach can improve performance based on the fundamental equations governing EHD, showing that absolute force output increases with number of stages, that overall force density increases as devices are miniaturized (i.e., drift gap is decreased), and that average thrust efficiency of a multi-stage device is dominated by the inter-stage losses.

The analytical model guides the design of a device implementation, and then experiments are performed for validation. A laser microfabrication process and the folding-and lamination-based assembly technique it enables are discussed. Experimental measurements of the current-voltage and outlet air velocity-voltage relationships are presented for 1-, 2-, and 3-stage devices. Device lifetime measurements show that the UV-laser defined corona emission tips are robust to degradation in the discharge plasma. A three-stage device exhibits a peak areal thrust of about 15 N/m$^2$, a force density of about 2 kN/m$^3$, and a volumetric power density of about 4 kW/m$^3$, among the highest values ever measured for EHD devices in the literature.

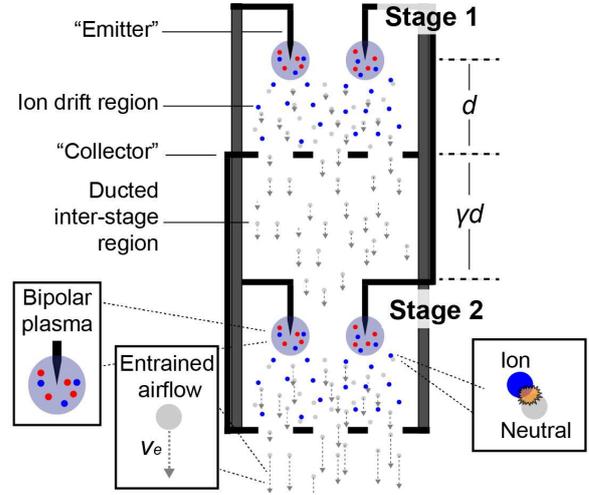

*Figure 1: The devices are composed of successive electrohydrodynamic acceleration stages, which each use corona discharge. Ions are ejected from the corona plasma region surrounding the emitters and drift in the applied electric field, colliding with neutral air molecules and imparting momentum. The accelerated air column is ducted as it travels between stages to reduce losses.*

**Performance Model**

Electrohydrodynamic force is produced by the momentum-transferring collisions of charged particles, drifting under the influence of an applied electric potential, with neutral fluid molecules (Fig. 1). When the neutral fluid is air this effect is sometimes known as electroaerodynamic (EAD) force. A one-dimensional derivation for the total force acting on a volume of ions relates the ion travel distance, $d$, the ion current, $I$, and the ion mobility, $\mu$. Assuming that momentum transferring collisions are frequent, this Coulomb force is approximately equal to the force imparted to the neutral air, with deviation captured by a loss parameter $\beta_1$. As corona discharge-based EHD is a space charge limited effect governed by the Mott-Gurney limit, the force can then be related directly to the drift field magnitude $E$ [3]:

$$F_0 = \beta_1 qE = \beta_1 \int_0^d \rho E\, dV = \beta_1 = \left(\frac{9}{8}\right)\beta_1 \epsilon_0\, A\, E^2 \quad (1)$$

indicating that areal force density depends primarily on applied electric field. Prior work has shown that areal force densities over 1N/m² can be maintained down to sub-millimeter ion drift distance devices, which have a lower operating voltage than higher gap devices [4].

Using (1), the absolute force output $F$ and the volumetric force density $\mathbb{F}$ in a multi-stage device, the latter neglecting the thickness of the emitter and collector electrodes and the device area outside the active region, is then given by:

$$F_n = n\, \beta_2\, F_0 \quad (2)$$

$$\mathbb{F} = \frac{n}{(n + [n-1]\gamma)\, d}\, \beta_2\, F_0 \quad (3)$$

where $n$ is the number of stages, $\gamma$ is the inter-stage spacing as a fraction of the drift gap $d$ (see Fig. 1), and $\beta_2$ is a measure of the average inter-stage loss (i.e., it is less than one). It is clear that decreasing the drift gap while maintaining a constant electric field increases the overall device force density. Force density decreases, however, with increasing number of acceleration stages due to the necessary inter-stage spacing and the introduction of the inter-stage loss factor, with the limiting case given by:

$$\lim_{n\to\infty} \frac{\beta_2 n}{n + (n-1)\gamma} = \frac{\beta_2}{1+\gamma} \quad (4)$$

Electrostatic interference between the electrodes of neighboring stages could influence the system performance, for example by decreasing the electric field magnitude at the emitter and therefore increasing the corona ignition voltage. Gilmore et al. show that an inter-stage spacing of $2d$, where $d$ is the electrode gap, is sufficient shielding; this corresponds to a $\gamma$ of two [5].

Another important performance parameter to consider is the force efficiency, $\eta$, in N/W. Using (1), the force efficiency of an electrohydrodynamic device, given the simplifying assumption that $E_{drift} = V/d$, is:

$$\eta = \frac{F}{P} = \frac{\beta_1 \rho E dA}{V(jA)} = \frac{\beta_1}{\mu E + v_{in}} \quad (5)$$

where $v_{in}$ is the neutral air velocity into the device. The inlet air velocity term is typically neglected, because for a static single stage device it is equal to approximately zero and for a moving (e.g., a flying robot thruster) device, the ion drift velocity, given by $v_{ion} = \mu E$, is typically on the order of hundreds of meters per second while the drag limited forward flight speed is only on the order of a few meters per second. The number of stages in a multi-stage device, however, will directly trade off with the achievable thrust efficiency, as successive stages will have higher inlet air velocities. Using simple momentum theory to find the outlet velocity of a stage given the EHD force and inlet velocity through the relation $F = \frac{1}{2}\rho A(v_e^2 - v_{in}^2)$ where $\rho$ is the air density, for a multi-stage device the average thrust efficiency can then be calculated as:

$$\eta_{ave} = \frac{\beta_2}{n}\sum_{i=1}^{n} \eta_i = \frac{\beta_1 \beta_2}{n}\sum_{i=1}^{n}\left(\mu E + \sqrt{\frac{F_{i-1}}{0.5\rho A}}\right)^{-1} \quad (6)$$

where $F_i$ is the multi-stage force given by (2) and $\eta_i$ is the single stage thrust efficiency given by (5). This equation shows that the average thrust efficiency will decrease with an increasing number of stages. Substitution of realistic values, however, shows that the inter-stage loss $\beta_2$ is the dominant parameter (see Table 1).

*Table 1: Decrease in average thrust efficiency from the single-stage case with increasing stage count and change in inter-stage loss, at a drift field of 1MV/m and $\beta_1 = 1$.*

|  | **$\Delta\eta$: $1 - \eta_{ave}/\eta_1$** | | |
| --- | --- | --- | --- |
|  | *3 stages* | *10 stages* | *20 stages* |
| $\beta_2 = 1$ | 1.4% | 3.3% | 4.8% |
| $\beta_2 = 0.8$ | 21.1% | 22.6% | 23.8% |

The analytical model shows that multi-stage device performance is largely controlled by the ion drift distance, the inter-stage distance, the number of stages, and the inter-stage losses. Minimizing the drift gap size and the inter-stage loss is most critical to a high force output, efficient design.

## DEVICE FABRICATION AND ASSEMBLY

Components are laser-microfabricated using a DPSS Samurai UV marking system. Unlike prior work demonstrating laser-microfabricated EHD devices [6], devices are assembled using a process where components are laminated into a three-dimensional structure using a combination of folding and a hot press adhesive (Fig. 2). Stainless steel is chosen for the active electrodes due to its mechanical strength and its theoretical resistance to oxidation in the corona plasma. Each stage has five radially arranged emitters ($\theta_{tip} = 2.5°$, nominally). Dielectric spacers are fabricated from polyetherimide (PEI), a high-k polymer.

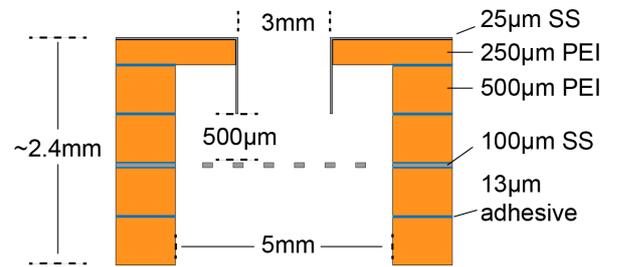

*Figure 2: Cross-section of one assembled stage. The PEI film laminated with dry mount adhesive and the 18-8 stainless steel (SS) shim stock are cut using a 355-nm DPSS laser.*

The entrained airflow is ducted using the PEI dielectric spacers in an effort to minimize loss of force in the desired axis (i.e., to yield a large value for $\beta_1$ and $\beta_2$). Perforations in the stainless-steel emitter and collector electrodes allow multiple EHD stages to be formed from a single folded electrode set (Fig. 3), while perforations at the base of the emitter tips allow them to be positioned towards the collector grids using a milled aluminum stamp and die set during assembly.

The components are layered in the die until reaching the first emitter stage then laminated using a heated press, which applies pressure and conducts heat through the base of the stamp. The subsequent collector and emitter stages are folded using the perforated stage connectors (Fig. 3 inset) and the process repeats. A three-stage folded device is shown in Figure 4.

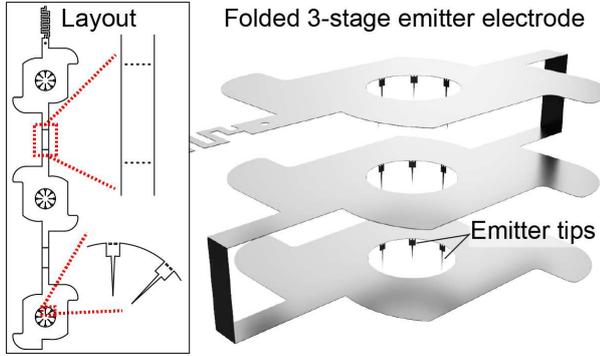

*Figure 3: Stainless steel electrodes (render of emitter electrode shown) have perforations (inset) to allow for out of plane folding into the desired three-dimensional structure during assembly.*

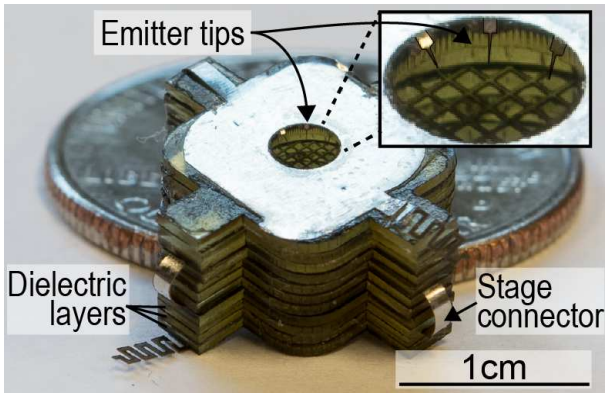

*Figure 4: Image of a folded three-stage device with U.S. quarter for scale.*

## EXPERIMENTAL RESULTS

A hot wire anemometer fixed under the EHD device outlet measures air velocity synchronized with applied voltage and measured current (both from an external high voltage power supply). Output force and power can then be estimated using these air velocity measurements and simple momentum theory. A schematic of the test setup is shown in Figure 5.

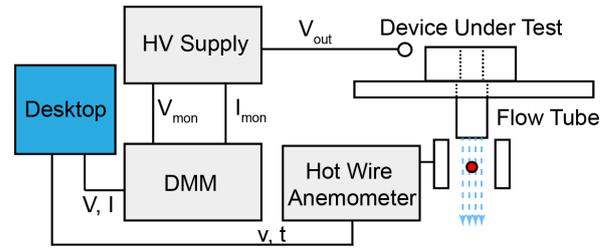

*Figure 5: The experimental setup for voltage-current-outlet air velocity measurements.*

Figure 6 shows IV curves for one-, two-, and three-stage devices. The corona discharge onset voltage is calculated as approximately 1600V, though this value varied by as much as $\pm200$ depending on fabrication and assembly variability.

Near linear scaling of the output current with number of stages indicates that the inter-stage spacing is sufficient and that the folded electrode design is successful.

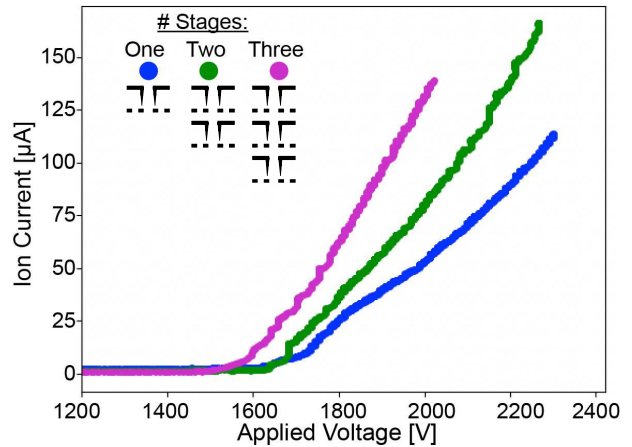

*Figure 6: Measured current as a function of applied voltage between the emitter and collector electrodes for one-, two-, and three-stage devices. Data is from single experiments with previously unused devices.*

Figure 7 shows the outlet air velocity as a function of applied voltage, measured simultaneously with the current in Figure 6. Based on (2), multi-stage devices are expected to have outlet air velocities $\propto \sqrt{n_{stages}}\, v_0$, where $v_0$ is the outlet velocity of a single stage device. Limited low-velocity measurement resolution prevented more accurate curve determination, but the measured output velocity follows the expected trend. Peak force density occurs at an efficiency of approximately 1.1 mN/W.

Prior work using silicon lithographically defined emission points showed degradation of the emitter electrode material began to occur immediately after corona ignition, with a drop in measured ion current of over 30% after 100 seconds of emission time [7]. Figure 8 shows that measured ion current from the stainless-steel laser microfabricated emitters used in this work decreases by only approximately 5% after 100 seconds. Change in the measured outlet air velocity was within the standard measurement error range. This indicates that the stainless-steel emitters are relatively robust to mechanical degradation in the plasma, but longer lifetime tests are warranted for use in high duty-cycle applications like electronics cooling.

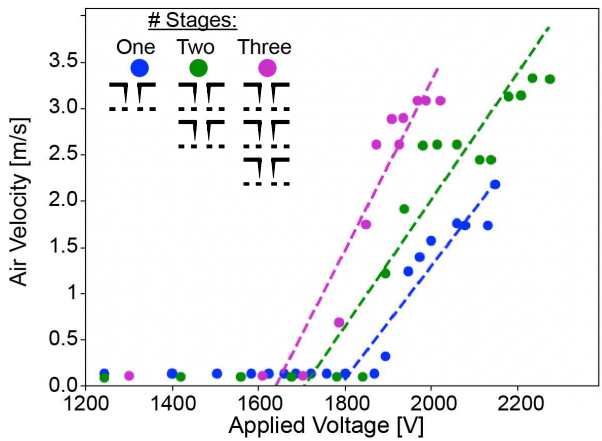

*Figure 7: Air velocity as a function of applied voltage between the emitter and collector electrodes for one-, two-, and three-stage devices. Linear regression lines for points beyond the inception voltage are shown.*

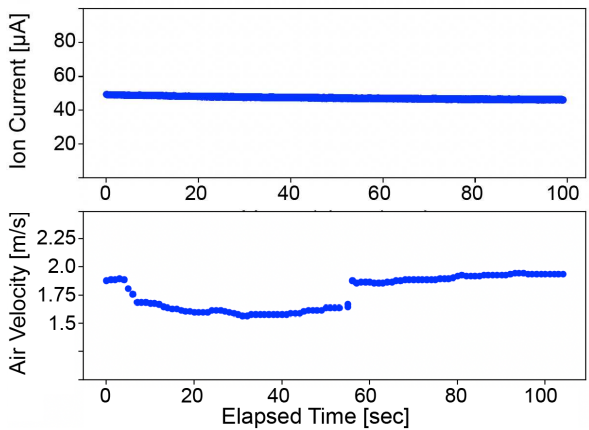

*Figure 8: Measured emission current as a function of time at a constant input voltage. Experiment performed with a single-stage device at 2000V.*

## CONCLUSION

EHD devices composed of multiple small ion drift gap stages stacked in series are high power density, solid state, silent actuators. This work explains the design considerations for a multi-stage device and demonstrates the first ever multi-stage, sub-millimeter drift gap actuator produced using laminated laser microfabricated electrodes.

Future work includes more rigorous investigation of how ducting and device geometry affect the inter-stage loss factor between subsequent acceleration stages, how device performance follows theory at high stage counts, and methods for minimizing device mass in order to increase specific power. Further, while (1) and (5) indicate that EHD force and efficiency can be set by the drift field and is independent of drift gap, in practice there are significant challenges associated with device miniaturization which cause both deviations from ideality (i.e., a small $\beta_1$ value) and inability to operate along the full theoretical range [8]. Future designs must overcome this challenge in order to push the boundaries of force density.

## ACKNOWLEDGEMENTS

This research was supported by the Intelligence Community Postdoctoral Research Fellowship Program, administered by Oak Ridge Institute for Science and Education through an interagency agreement between the U.S. DoE and the ODNI.

## CONTACT
Daniel Drew; dsdrew@stanford.edu